# Spin pumping in d-wave superconductor/ferromagnet hybrids

S. J. Carreira[1,*], D. Sanchez-Manzano[1,2], M.-W. Yoo[1], K. Seurre[1], V. Rouco[1], A. Sander[1], J. Santamaría[2], A. Anane[1], and J. E. Villegas[1]

[1]Unité Mixte de Physique, CNRS, Thales, Université Paris-Saclay, 91767, Palaiseau, France.

[2]Grupo de Física de Materiales Complejos (GFMC). Dpto. de Física de Materiales. Facultad de Ciencias Físicas, UCM, Plaza Ciencias, 1 28040, Madrid, Spain.

Spin-pumping across ferromagnet/superconductor (F/S) interfaces has attracted much attention lately. Yet the focus has been mainly on s-wave superconductors-based systems whereas (high-temperature) d-wave superconductors such as $YBa_2Cu_3O_{7-d}$ (YBCO) have received scarce attention despite their fundamental and technological interest. Here we use wideband ferromagnetic resonance to study spin-pumping effects in bilayers that combine a soft metallic $Ni_{80}Fe_{20}$ (Py) ferromagnet and YBCO. We evaluate the spin conductance in YBCO by analyzing the magnetization dynamics in Py. We find that the Gilbert damping exhibits a drastic drop as the heterostructures are cooled across the normal-superconducting transition and then, depending on the S/F interface morphology, either stays constant or shows a strong upturn. This unique behavior is explained considering quasiparticle density of states at the YBCO surface, and is a direct consequence of zero-gap nodes for particular directions in the momentum space. Besides showing the fingerprint of d-wave superconductivity in spin-pumping, our results demonstrate the potential of high-temperature superconductors for fine tuning of the magnetization dynamics in ferromagnets using k-space degrees of freedom of d-wave/F interfaces.

*Corresponding author: santiago.carreira@cnrs-thales.fr

**Introduction**

Spin injection into superconductors constitutes a very active research topic within the nascent field of superconducting spintronics, aiming at expanding spintronic functionalities by exploiting the dissipationless electron transport and quantum coherence characteristic of superconductivity [1–5].

Theory and experiments have shown that spin currents can flow into s-wave superconductors carried by equal-spin triplet Cooper pairs [1,2,6–9] or by superconducting quasiparticles [10,11], whose lifetime can exceed those of spin-polarized electrons in the normal state [12–16]. Spin-polarized quasiparticles can be efficiently injected into the superconductor (S) using an adjacent ferromagnet (F) by applying across the S/F interface a bias voltage that exceeds the superconducting gap [10,17]. This mechanism has been extensively explored in transport experiments with spin valves [13,18–21]. Another mechanism for inducing a non-equilibrium spin accumulation in superconductors is spin-pumping [22] using the resonant excitation of the ferromagnet's magnetization [23,24] as source of pure spin current. In these ferromagnetic resonance (FMR) experiments, the superconductor's efficiency as a spin-sink is evaluated via spin hall effect [25] or microwave absorption measurements [8,25–29], by monitoring the evolution of the resonant peak's linewidth across the superconducting transition. The assumption is that the changes of the magnetic damping (which lead to a narrowing/broadening of the resonance linewidth [23,24]) reflect variations in the spin relaxation rate when the superconducting gap opens, because this alters both the spin transmission across the superconductor/ferromagnet interface and the relaxation mechanisms within the superconductor. Pioneering experiments performed on $Ni_{80}Fe_{20}/Nb$ (Py/Nb) bilayers have found that the opening of the superconducting gap induces an abrupt FMR linewidth narrowing when temperature is swept across the superconducting transition [26]. This was explained by considering that the opening of the superconducting gap leads to a vanishing density of states at the Fermi level, thereby hindering the transmission of spin polarized electrons across the interface. More recent work on GdN (F) / NbN (S) multilayers has found a different behavior, in which the Gilbert damping initially peaks across the superconducting transition, and diminishes below the normal-state value upon further temperature decrease [30]. That behavior was associated to the presence of spin-orbit scattering at the interface [31]. In contrast to the two examples mentioned above, studies carried out on Py/Nb multilayers with an adjacent strong spin-orbit coupling metal (Pt) found a steady broadening of the linewidth below $T_C$, which was interpreted in terms of enhanced spin transport across the superconductor due to the generation of equal-spin triplet superconductivity [7,8]. Adding a new piece to the puzzle, a recent theory shows that, if the superconducting gap is suppressed near the S/F interface, the presence of quasiparticle surface states can also produce an enhancement of spin transport into the superconductor below $T_C$ [32]. The strikingly wide variety of observed behaviors illustrates the complexity of the underlying physics, the importance of the interfacial properties, and the fact that the conditions for predominance and interplay of the different proposed scenarios (quasiparticles and triplet superconductivity) is far from being fully understood. Beyond raising these fundamental questions, it is interesting that the experimental investigations have evidenced that superconductivity may be exploited for tuning magnetization dynamics.

The experiments discussed so far are based on conventional (low-$T_c$) s-wave superconductors, which present an isotropic superconducting gap. In contrast, in unconventional (high-$T_c$) d-wave ones the gap is suppressed along particular directions in the momentum space, and there exists a π superconducting-phase shift between d-wave lobes [33–35]. While spin diffusion effects in d-wave superconductors have been discussed in the context of electrical measurements [36–41], to our knowledge spin-pumping and the effects of the onset of superconducting pairing on the spin-sink behavior of d-wave cuprates remain unexplored. Notice that, at variance to s-wave superconductors, the presence of zero-gap nodes may provide channels for injection of spin-polarized electrons, even in the superconducting state. Consequently, the effects of superconductivity on spin-pumping and magnetization dynamics are expectedly different in the case of s-wave superconductors. Here we experimentally investigate this issue using c-axis YBCO/Py heterostructures with different interface structure. In all cases, we observe an abrupt linewidth narrowing across the superconducting transition, similar to that observed in Py/Nb s-wave system [26], which suggests that, right below the critical temperature, the opening of the d-wave gap significantly suppress spin-pumping. However, upon further temperature decrease, the behavior of the linewidth depends on the YBCO surface morphology. For the smoother YBCO films, we observe no further evolution of the linewidth. However, in the presence of a faceted YBCO surfaces, the linewidth monotonically widens as the temperature is decreased below $T_c$. This behavior can be explained considering the interfacial density of quasiparticle states, which depends on the YBCO surface morphology due to the anisotropic character of d-wave superconductivity. These results thereby provide a fingerprint of d-wave superconductivity in in the physics of spin-pumping. At the same time, they underline the need of a theoretical framework that specifically addresses the role of the mechanisms at play (quasiparticle density of states [32,42], changes in the spin-imbalance relaxation [43] and dynamic generation of triplet pairs [44,45]) in the context of d-wave superconductivity. Finally, this work demonstrates the potential of high-temperature superconductors for manipulating the magnetization dynamics of metallic ferromagnets, in a way that could be engineered by choosing the orientation of the d-wave/F interface.

**Experimental**

We have studied different multilayers, namely c-axis $YBa_2Cu_3O_7$ (30 nm)/$Ni_{80}Fe_{20}$ (15 nm)/Al (3 nm) grown on (001) $SrTiO_3$ (one sample) and on (001) $NdGaO_3$ (two samples) – respectively referred to as STO//S/F, NGO//S/F #1 and NGO//S/F #2 – and YBa$_2$Cu$_3$O$_7$ (30 nm)/Au (5 nm)/Ni$_{80}$Fe$_{20}$ (15 nm)/Al (3 nm) on STO –referred to as STO//S/Au/F. The YBCO films were grown by pulsed laser deposition (PLD) using an excimer laser ($\lambda$ = 305 nm) at a temperature of 700 °C and oxygen pressure of 0.36 mbar. Optimum oxygenation was ensured by raising the O$_2$ pressure to 760 mbar during cooldown. Where applicable, the Au interlayer (aimed at preventing and assessing the impact of eventual redox reactions between YBCO and Py) was subsequently grown in-situ by PLD, at room temperature and in pure Ar atmosphere. Under these growth conditions, the onset of the superconducting transition determined by resistivity measurements is typically around Tc ~ 85 K, regardless of the substrate and presence of an Au interlayer.

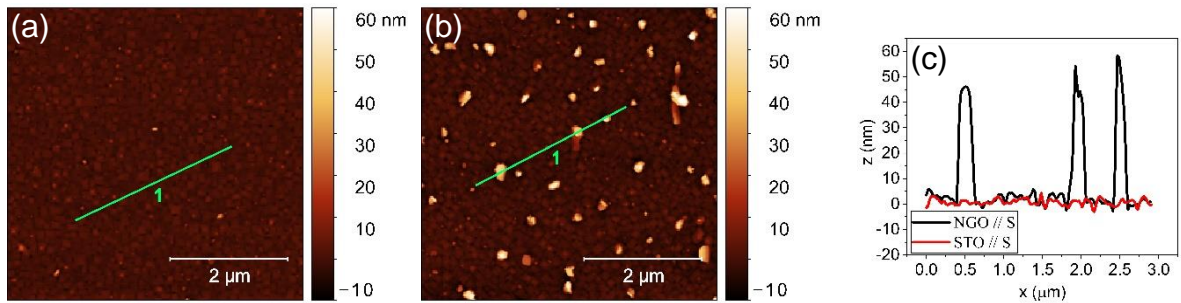

FIG. 1. AFM images measured on a 5x5 μm$^2$ area of a YBCO thin film grown on (a) STO (001) and (b) NGO (001). The height profile shown in (c) was measured along the oblique line 3 μm long indicated in (a) and (b) respectively.

The structural properties of the as-grown YBCO films were studied by high-angle X-ray diffraction, which confirmed c-axis (001) epitaxial growth on both substrates STO and NGO, as well as the absence of parasitic phases (see Fig. S1 in Supplemental Material). However, we found that the YBCO's surface morphology is different depending of the substrate. Atomic Force Microscopy (AFM) images displayed in Fig. 1 show that YBCO on STO Fig. 1 (a) presents a relatively smooth surface (rms roughness ~ 2 nm), while YBCO on

NGO [Fig. 1 (b)] presents a high density of ~ 50 nm tall crystallites [see profile in Fig. 1 (c)] on top of an otherwise similar background topography. The Py layer and Al capping (aimed at preventing Py surface oxidation) were subsequently deposited on the YBCO *ex-situ*, using rf-sputtering in pure Ar atmosphere at room temperature, without breaking vacuum between each layer deposition. Control samples consisting of single Py films grown on both $SrTiO_3$ and $NdGaO_3$ (labeled as STO//F, STO//Au/F, NGO//F #1 and NGO//F #2) were studied. The samples' size is in all cases 5×5 mm$^2$.

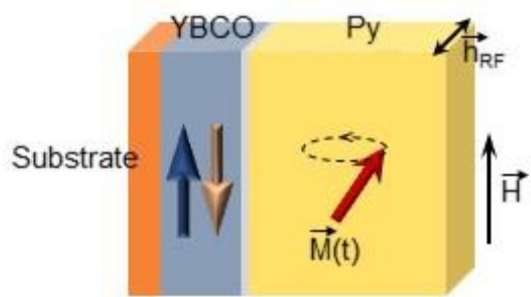

FIG. 2. Sketch of the multilayer structure and experimental geometry for the FMR experiments.

The experimental geometry considered for the FMR experiments is sketched in Fig. 2. A DC magnetic field H is applied parallel to the sample plane in order to saturate the magnetization of the Py, whose precession is excited by applying and a radiofrequency (RF) magnetic field $h_{RF}$ perpendicular to the DC field, using a coplanar waveguide. A magnetic field modulation at low frequency (< 2 kHz) is used to measure the derivative of the absorbed power dP/dH with respect to the DC magnetic field H, as this is swept around the resonance field $H_{res}$ where the dynamical susceptibility peaks. A typical measurement is shown in Fig. 3 (a). This type of measurements were done for a number of fixed frequencies in the range 4 GHz < f < 40 GHz. For each frequency, the peak-to-peak linewidth $\Delta H_{pp}$ and the resonance field $H_{res}$ were determined by fitting the dP/dH *vs.* the applied field H to the derivative of a Lorentzian function, as is shown in the example of Fig. 3 (a). This allows extracting the values of the resonance field $H_{res}$ and linewidth $\Delta H_{pp}$ versus the frequency, which are shown in Figs. 3 (b) and (c) for the example in (a). The relationship between the resonant microwave frequency f and field $H_{res}$ is given by the Kittel formula [46] which, neglecting the small magnetic anisotropy of Py, is

$$f = \gamma \mu_0 \sqrt{H_{res}(H_{res} + M_{eff})} \qquad (\text{Eq. 1})$$

where γ is the gyromagnetic factor and $M_{eff}$ is the effective magnetization. The linewidth is well described by the linear expression [24],

$$\mu_0 \Delta H_{pp} = \frac{2\alpha f}{\sqrt{3}\gamma} + \mu_0 \Delta H_0 \qquad (\text{Eq. 2})$$

where $\mu_0 \Delta H_0$ is the frequency-independent contribution or inhomogeneous broadening and α is the Gilbert damping factor. Similarly as in the example shown in Fig. 3, the data for all the studied samples is well described by Eq. 1 and Eq. 2. This allowed us to obtain the temperature dependent α and $\mu_0 \Delta H_0$ for the series of samples, with error bars calculated from the linear regression of the fits. Notice that, based on the linear behavior observed in $\mu_0 \Delta H_{pp}$ *vs.* f for a broadband frequency range in all of the studied samples, we consider that the 2-magnon scattering can be ruled out as a dominant relaxation mechanism [47] in all of them.

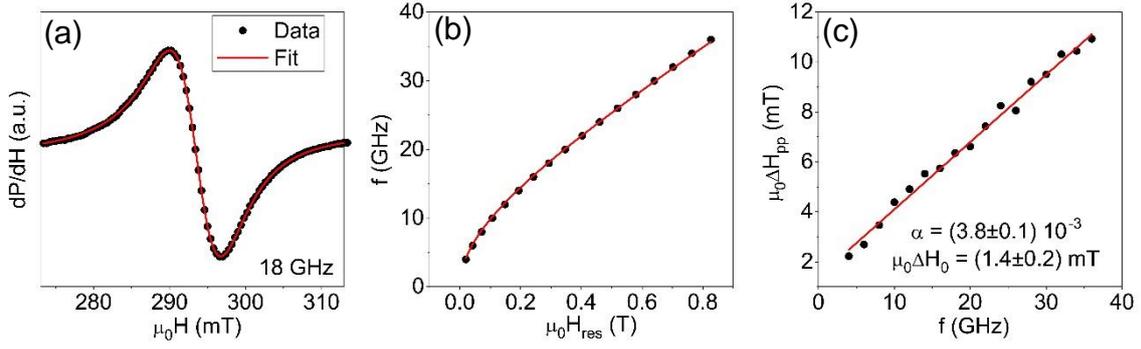

FIG. 3. Typical (a) FMR absorption spectrum and fit, (b) f vs $\mu_0 H_{res}$ and (c) $\mu_0 \Delta H_{pp}$ vs f obtained for the sample STO//S/Au/F at 30 K. The fits in (b) and (c) follows the FMR equations (1) and (2).

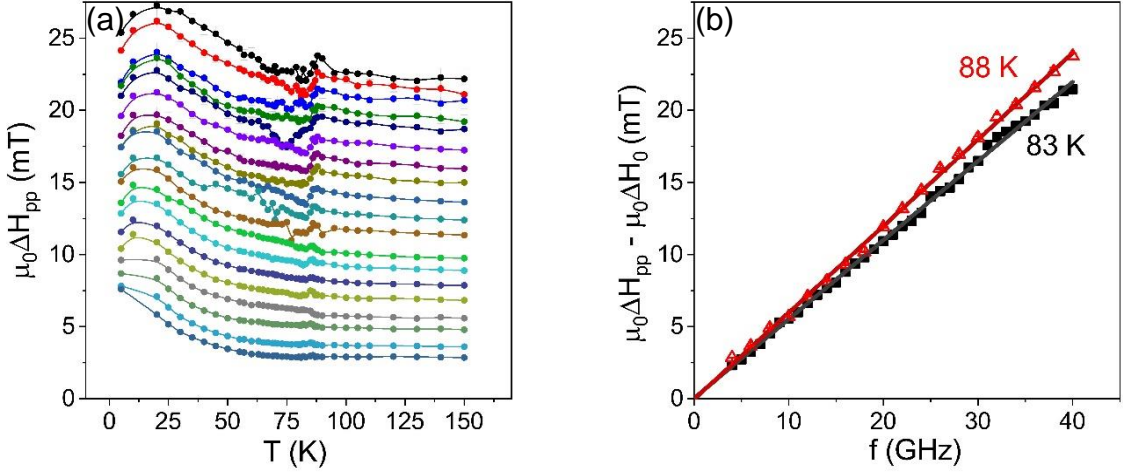

FIG. 4. (a) Temperature dependence of the FMR linewidth, $\mu_0\Delta H_{pp}$, measured at all frequencies from 4 GHz to 40 GHz in steps of 2 GHz for the sample NGO//S/F #2. (b) $\mu_0\Delta H_{pp} - \mu_0\Delta H_0$ vs f for the sample NGO//S/F #2 obtained at temperatures just above (88 K) and below (83 K) the superconducting critical temperature of the YBCO. The straight lines correspond to linear fits of the data points.

**Results**

Fig. 4 (a) shows, as an example, a typical series of the temperature-dependent FMR linewidth $\mu_0\Delta H_{pp}$ measured for different frequencies, which corresponds to a NGO//F/S sample. The background trend −a steady linewidth broadening with decreasing temperature, with a drop below ~ 20 K for the measurements at highest frequencies− is similar to that of the NGO//F reference samples (see Fig. S2(a) in Supplemental Materials) and to the behavior observed in earlier FMR experiments on single Py thin films [47–50]. On top of that background, we observe another feature, a "kink" around T ~ 85 K, which is not present in the reference samples and, as discussed below, is related to superconductivity. However, the fact that $\mu_0\Delta H_{pp}$ results from the addition of the (frequency independent) inhomogeneous broadening and the (frequency dependent) magnetic damping, makes such feature evident only for f > 18 GHz.

This feature indeed corresponds to a drop of the damping factor α across the superconducting transition, as evidenced in Fig. 4 (b) where the linewidth (after subtraction of the frequency-independent broadening $\mu_0 \Delta H_0$) is plotted as a function of the frequency. One can see that the damping (slope of the straight lines) is different above (88 K) and below (83 K) the superconducting transition of YBCO.

The above example makes it evident that broadband measurements are crucial to finely quantify the linewidth changes across the superconducting transition, and to univocally ascribe them to a variation of the damping factor. Thus, in what follows, we will compare samples based on the temperature dependence of the damping coefficient α(T), which can be obtained together with the temperature dependent inhomogeneous broadening $\mu_0\Delta H_0(T)$ by applying the analysis described above to a series of $\mu_0\Delta H_{pp}$ vs f measured at different temperatures.

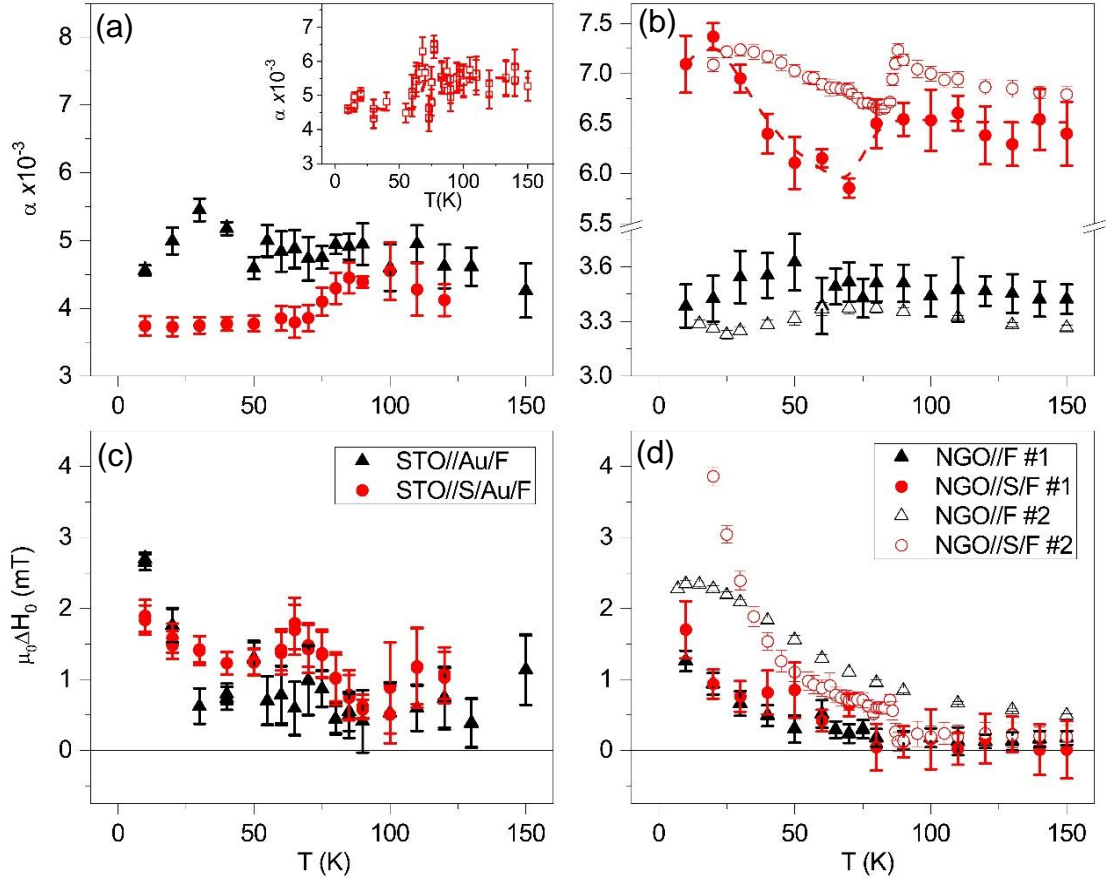

FIG. 5. Temperature dependence of the [(a) and (b)] magnetic damping $\alpha$ and [(c) and (d)] inhomogeneous broadening $\mu_0\Delta H_0$ for the samples STO//S/Au/F and STO//Au/F [(a) and (c)] and NGO//S/F and NGO//F [(b) and (d)]. In (b) and (d) we plot the results obtained for two samples with the same nominal composition, #1 (filled symbols) and #2 (open symbols). Data in circles corresponds to the samples with YBCO as a bottom layer and the control samples without YBCO are denoted with triangles. The inset in (a) shows $\alpha$ vs T for the sample STO//S/F. The dash lines are guides to the eye.

In Fig. 5 (a) we show $\alpha(T)$ for superconducting multilayers STO//S/Au/F (red circles, main panel) and STO//S/F (inset), together with the data (black triangles) for a single Py film (sample STO//Au/F) used as reference. One can see that, when Py is combined with the superconductor, and regardless of the presence of an Au interlayer, $\alpha(T)$ drops by ~10-15% between 90 K and 70 K. Upon further temperature decrease $\alpha(T)$ stays nearly constant. That is, $\alpha$ drops across the superconducting transition, and remains constant thereafter. This contrasts with the behavior of the STO//Au/F sample used as reference (black dots), which shows no clear change of α around that temperature range. Notice also that the damping level α ~ 4.5 $10^{-3}$ in the temperature range in which the YBCO is in the normal state (T > 90 K) is comparable for the superconducting (STO//S/Au/F) and reference (STO//Au/F) samples. Fig. 5 (c) shows that $\mu_0\Delta H_0(T)$ behaves very similarly in the superconducting and reference samples. This implies that the presence of the YBCO does not create additional magnetic inhomogeneities in Py, and unambiguously demonstrates a decrease of the Gilbert damping across the superconducting transition. This effect can also be observed in the NGO//S/F #1 and #2 bilayers [see Fig. 5 (b)] for which α(T) shows a ~ 10% drop across the superconducting transition (red circles) not observed in the reference NGO//F sample (black triangles). As was pointed out for the STO substrate, the inhomogeneous broadening is not significantly affected by the presence of the YBCO layer, see Fig. 5 (d). However, there are two main differences when comparing samples grown on STO and on NGO. First, for NGO//S/F the damping level α ~ 6.5 $10^{-3}$ in the normal-state (T > 90 K) is significantly higher than for the reference sample NGO//F [Fig. 5 (d)]. Second, for NGO//S/F the magnetic dumping α(T) does not remain constant below the superconducting transition, but shows instead an upturn with decreasing temperature.

**Discussion**

The central observation is that the magnetic damping α(T) of Py in YBCO/Py heterostructures drops across the YBCO superconducting transition and that, upon further temperature decrease, α(T) either stays constant or shows an upturn depending on the substrate (STO or NGO) on which the heterostructures are grown. The initial drop across the transition is reminiscent of that observed in earlier experiments with s-wave superconductors [26], which was explained based on the idea that, as the superconducting gap in the electronic density of states opens [48], the decrease of electrons states at the Fermi level impedes spin injection. Such blocking effect strengthens as temperature is lowered further from T_C, because this makes the superconducting gap widen and the quasiparticle population diminish [48]. While such effect is consistent with the behavior of α(T) for heterostructures grown on STO, it cannot fully account for the behavior of the samples grown on NGO: these show an upturn of the damping factor, which at low temperature reaches values higher than those observed above T_C [Fig. 5 (b)]. A similar enhancement of spin-pumping in the superconducting phase was observed in S/F interfaces [7,8] in the presence of a heavy metal (Pt), and was explained by the generation of equal-spin triplet pairs. However, in the present experiments we have no arguments nor evidence to support such scenario. Instead, we have considered a different situation recently studied theoretically [32], in which an enhancement of spin-pumping in the superconducting phase is explained the presence of a quasiparticle states (Andreev bound states) at the interface with the F. In Ref. [32] s-wave superconductors were considered, for which the emergence of Andreev bound states stems from the interfacial suppression of the superconducting gap due to inverse proximity effect. However, in the case of d-wave superconductors quasiparticle (Andreev) surface bound states appear *intrinsically*, due to the existence of zero-gap nodes along particular k-space directions [49]. As we detail below, the quasiparticle density depends on the interface orientation. This provides a possible scenario to explain the distinct behaviors of samples grown in STO and NGO based on their different surface topography.

Follwing [32], the spin-pumping into the S depends on the surface density of quasiparticle states: the larger the density of states, the larger the spin injection efficiency. Extending the full calculations existing for s-wave superconductros [32] to the case of d-wave is out of the present work's scope. However, a qualitative explanation for experimental results is at reach by considering the density of quasiparticle states at d-wave/normal metal interfaces with finite transparency. Following [50], the normalized density of quasiparticle states is:

$$\rho_{S_0}/\rho_N (E) = \frac{1-(\sigma_N-1)^2|\Gamma_+\Gamma_-|^2}{|1+(\sigma_N-1)\Gamma_+\Gamma_-\exp(i\phi_- - i\phi_+)|^2} \qquad \text{(Eq. 3)}$$

where $\rho_N$ is the normal-state electron density of states, $\sigma_N = \frac{1}{1+Z^2}$ with Z the barrier strength at the interface, $\Gamma_\pm = \frac{E - \sqrt{E^2 - |\Delta_\pm|^2}}{|\Delta_\pm|}$ with E the quasiparticle energy with respect to the Fermi level, and $\phi_+$ (respectively $\phi_-$) is the effective phase of the anisotropic pair potential $\Delta_+ (\Delta_-)$. Temperature effects in the quasiparticle population can be taken into account by considering the gap amplitude $\Delta(T) = \Delta_0 \tanh(b\sqrt{\frac{T_C}{T} - 1})$ and by convoluting $\rho_{S_0}/\rho_N (E)$ with the derivative of the Fermi-Dirac distribution $f_{FD}(E,T)$ [51],

$$\rho_S/\rho_N (E,T) = \int \rho_{S_0}/\rho_N (E') \frac{\partial f_{FD}}{\partial E}(E - E', T) \, dE' \qquad \text{(Eq. 4)}$$

Calculations of the normalized density of states $\rho_S/\rho_N (E,T)$ for interfaces facing a d-wave gap lobe ($\alpha_g = 0$) and facing a gap node ($\alpha_g = \pi/4$) are shown in Fig. 6 (a) and (c) respectively, considering a moderate interface transparency Z = 2.5 (Fig. S3 of the Supplemental Materials demonstrates that, except for very transparent interfaces Z<1, the effects discussed thereafter are qualitatively similar for any Z). The different behaviors in Fig. 6(a) and 6(c) result from the anisotropic nature of the density of states at the YBCO surface. For a $\alpha_g = 0$ surface, we observe at low energies ($E < \Delta$) that the opening of the superconducting gap leads to a fast reduction of the density of states upon decreasing temperature, similarly as in s-wave superconductors. On the contrary, for the $\alpha_g = \pi/4$ case [Fig 6 (c)] we observe the emergence of Andreev bound states around $E = 0$, whose population gradually increases upon decreasing temperature, leading to a peak in the density of states. In our experiments, the microwave energy $\hbar f \ll \Delta$, and thus the relevant quantity is the density of states near the Fermi level ($E \sim 0$) [32]. This is shown in Fig. 6 (b) for the two cases $\alpha_g = 0$ and $\alpha_g = \pi/4$.

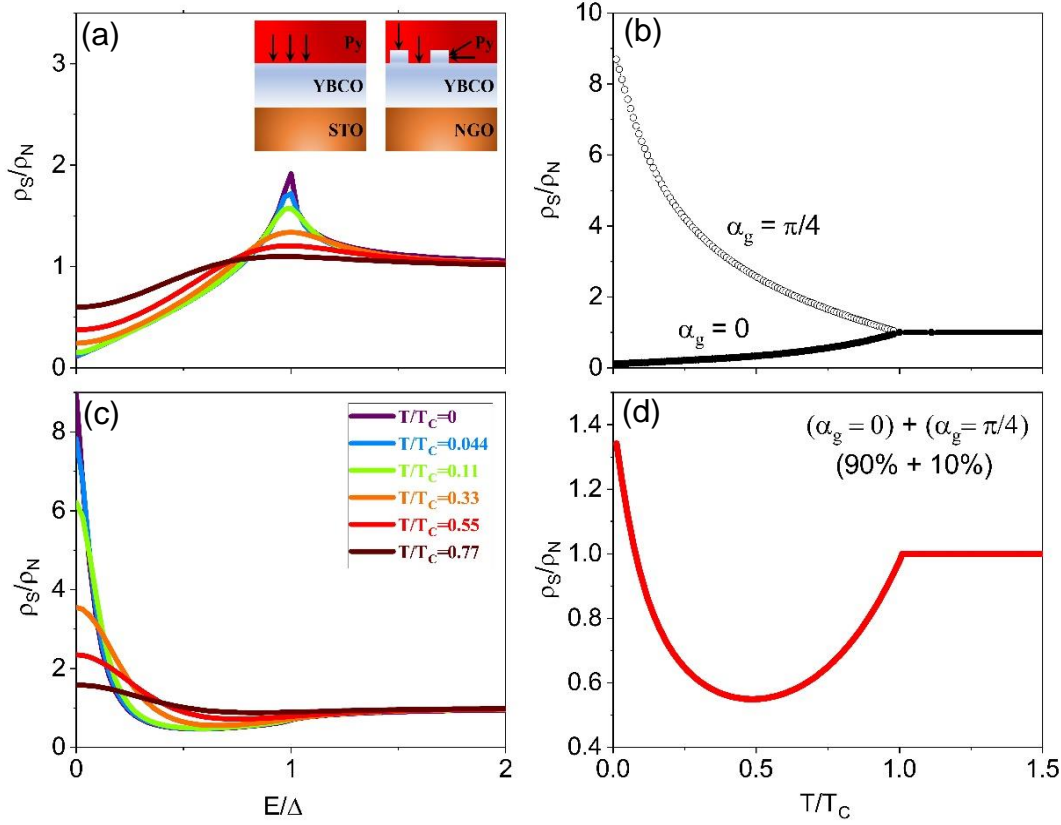

FIG. 6. Calculated density of states for an interface (a) facing a d-wave gap lobe $\alpha_g = 0$ and (b) facing at d-wave gap node $\alpha_g = \pi/4$ for different temperatures. The sketches in (a) illustrate the possible directions for the spin injection according to the surface morphology. In (b) we show the temperature dependence of the density of states for quasiparticles injected along the $\alpha_g = 0$ and $\alpha_g = \pi/4$ directions and in (d) we plot the resulting density of states when 10% / 90% contributions of the $\alpha_g = 0$ and $\alpha_g = \pi/4$ are considered for the spin injection.

Based on the above, and considering the different topography of the STO and NGO samples, a possible interpretation for the different $\alpha(T)$ emerges. As sketched in the inset of Fig. 6 (a), in the case of STO the effects along the out of plane direction dominate, because of the smoother S/F interface. In this situation, the density of quasiparticle surface states is as in Fig. 6 (a) [52] and, as was observed for s-wave supercoducters [26], we expect that $\alpha(T)$ decays across the superconducting transition, in agreement with our experimental findings [Fig. 5 (a)]. However, for samples grown on NGO the presence of crystallites at the surface allows spin pumping into the YBCO basal (*ab*) plane [sketch in the inset of Fig. 6 (a)], which provides access to a larger density of zero-energy quasiparticle states. If we consider that this results in an effective density of states in which the contribution of directions presenting a large density of Andreev bound states weigths 10%, the calculated $\rho_S/\rho_N\,(E,T)$ [Fig 6 (d)] qualitatively reproduce the behavior of $\alpha(T)$ in the experiments [red in Fig. 5 (b)]: an abrupt drop across the transition, followed by an upturn upon further temperature decrease. A 10% weight of directions with large zero-energy quasiparticle density is reasonable for the samples grown on NGO considering the lateral area of the crystallites, which can be estimated from the AFM images. As discussed in the Supplemental Material, the ratio between the lateral surface area (normal to the *ab* plane) and the horizontal one (normal to the c-axis) is between 1 % and 1.7 % depending on the criterion used for the estimate. Their contribution needs to be corrected due the large electronic anisotropy of YBCO, because the conductivity in the basal (*ab*) plane is up to 10 times larger than along the c-axis [53–55]. Thus, the 90%/10% contribution that allows reproducing the experimetnal results seem reasonable. We stress nevertheles that the discussed model aims a provinding a qualitative explanation of the observed behavior, and that the numerical estimates are made just to verify that the size of the effects are of the right order of magnitude.

Consistently with the scenario discussed above, we observe that for the NGO//S/F samples the normal-state damping is significantly higher than for the reference sample (see Fig. 5 (b) for T > 90 K), as Py contacts the YBCO not only on the c-axis surface but also on the more conducting basal (*ab*) plane. This results in a higher interfacial conductance than for the film grown on STO, which enhances the spin absorption and therefore the overall damping.

A final word concerning the impact of the Au interlayer. When the Au layer is deposited on YBCO, we observe no major effect on $\alpha(T)$, which indicates that its presence does not significantly change the interface transparency and is consistent with the fact the spin the diffusion length of Au ($\approx$ 50 nm at 10 K) [56] is larger than the Au layer thickness. In the control (non-superconducting) samples, the presence of an Au interlayer between Py and the insulating substrate enhances the magnetic damping, which reflects that Au is a more efficient spin-sink than the substrate.

In summary, we have found that in d-wave superconductor/ferromagnet YBCO/Py multilayers, the opening of the superconducting gap reduces the spin-sinking efficiency and results in a significant drop of the magnetic damping across the superconducting transition. However, upon further temperature decrease different behaviors are observed (either a plateau or an upturn), which can be associated with the YBCO's surface morphology. In particular, the low-temperature upturn can be explained by the large density of quasiparticle bound states characteristic of d-wave superconductivity. Our hypothesis is that those states are accessible via YBCO crystallites at the surface, that directly expose the YBCO *ab* plane to the interface with the ferromagnet. This suggests that spin-pumping into quasiparticle bound states could be further enhanced by engineering the YBCO surface, for example by growing YBCO in different crystallographic directions, or by creating vicinal surfaces. This, together with further theoretical developments -for instance an extension of Ref. [32] to the case of d-wave superconductors, possibly including other ingredients such as changes in the spin relaxation time in the superconducting state along different crystallographic directions [10,17]- would allow a more accurate quantitative analysis that would underpin the proposed scenario.

**Acknowledgements**

Work supported by ERC grant N° 647100 "SUSPINTRONICS", French ANR grant ANR-15-CE24-0008-01 "SUPERTRONICS", COST action "Nanocohybri" and EU H2020 Marie Curie Actions Nº 890964 "SUPERMAGNONICS".